\documentclass[12pt]{article}
\usepackage{amsfonts,a4,theorem}
\def\be{\begin{equation}}
\def\ee{\end{equation}}
\def\ie{{\it i.e.}}
\def\g{{\cal G}}
\def\title#1{\begin{center}\begin{large}{\bf #1}\end{large}\end{center}}
\def\author#1{\begin{center}#1\end{center}}
\def\address#1{\begin{center}{\sl #1}\end{center}}
\newcommand{\R}{{\ensuremath{\mathbb R}}} 
\def\borrar#1{}
\theorembodyfont{\rmfamily}
\newtheorem{lemma}{Lemma}
\begin{document}
\begin{flushright}
DAMTP 97--12, revised 28 May
\\
hep-th/9703019
\end{flushright}
\vspace{0.4cm}
\title{On the higher order generalizations of Poisson structures
\\[0.3cm]}

\author{J A de Azc\'arraga\dag 
\footnote{St. John's College Overseas Visiting Scholar.}
\footnote{On sabbatical (J.A.) leave and on leave of absence (J.C.P.B.)
from \ddag\ above.}, 
J M Izquierdo\ddag\ 
and J C P\'erez Bueno\dag \footnotemark[2]
\\[0.3cm]}

\address{\dag Department of Applied Mathematics and Theoretical Physics,
Silver St., Cambridge, CB3 9EW, UK
\\[0.2cm]}

\address{\ddag  Departamento de F\'{\i}sica Te\'orica and IFIC 
(Centro Mixto Univ. de Valencia-CSIC) E-46100 Burjassot (Valencia), Spain
\\[0.2cm]}

\begin{abstract}
The characterization of the Nambu-Poisson $n$-tensors as a subfamily 
of the generalized Poisson ones recently 
introduced 
(and here extended to the odd order case) is discussed.
The homology and cohomology complexes of both structures
are compared, and some physical considerations are made.
\end{abstract}

\section{Nambu-Poisson and Generalized Poisson structures}

{\it a) Nambu-Poisson structures}

The generalization of the Hamiltonian mechanics proposed by Nambu \cite{Na} 
more than twenty years ago has recently attracted a renewed attention, 
particularly since Takhtajan \cite{Ta} extended it further 
by introducing Poisson 
brackets (PB) involving an arbitrary number $n$ of functions, the case $n=3$ 
being Nambu's original proposal. His {\it Nambu-Poisson} 
(N-P) tensors provide an interesting generalization of the 
mathematical notion of {\it Poisson structure} (PS) on a manifold $M$ 
\cite{Lich}. A Nambu-Poisson 
structure is defined by a $n$-linear mapping $\{\cdot,\dots,\cdot\}:
{\cal F}(M)\times\mathop{\cdots}\limits^{n}\times{\cal F}(M)\to{\cal F}(M)$
which is:
a) completely antisymmetric; b) satisfies the Leibniz rule \ie,
$\{f_1,\dots,f_{n-1},gh\}=g\{f_1,\dots,f_{n-1},h\}+\{f_1,\dots,f_{n-1},g\}h$
and c) verifies the ($2n$--1)-point, ($n$+1)-terms 
{\it fundamental identity} (FI) 
\cite{Ta}
\be
\begin{array}{l}
\{f_1,\dots,f_{n-1},\{g_1,\dots,g_n\}\}=
\{\{f_1,\dots,f_{n-1},g_1\},g_2,\dots,g_n\}
\\[0.2cm]
+\{g_1,\{f_1,\dots,f_{n-1},g_2\},\dots,g_n\}
+\dots+\{g_1,\dots,g_{n-1},\{f_1,\dots,f_{n-1},g_n\}\}\quad.
\end{array}
\label{fi}
\ee
This relation may be understood as expressing that the time evolution 
for $(n-1)$ Hamiltonians $H_i\,,\,i=1,\dots,(n-1)$ given by
\be
\dot f=\{H_1,\dots,H_{n-1},f\}\quad,
\label{tev}
\ee
is a derivation of the $n$-N-P bracket. The case $n=3$ corresponds to Nambu's 
mechanics, although its associated five-point identity [eq. (\ref{fi}) 
for $n$=3], introduced by Sahoo and Valsakumar \cite{SaVa}, was not 
explicitly mentioned in his work.

The N-P bracket may be introduced through an antisymmetric contravariant 
tensor $\eta\in\wedge^n(M)$ or {\it multivector}, locally expressed by
\be
\eta={1\over n!}{\eta_{i_1\ldots i_n}} \partial^{i_1}
\wedge\ldots\wedge \partial^{i_n}\quad,\quad \partial^i
=\partial/{\partial x^i} \quad ,
\label{nvector}
\ee
by defining
\be
\{f_1,\dots,f_n\}=\eta(df_1,\dots,df_n)\quad.
\label{nbrdef}
\ee
Since (\ref{nvector}),(\ref{nbrdef}) automatically guarantee properties 
a), b) above, all that is required from $\eta$ is to satisfy the FI.
It is shown in \cite{Ta} that this is achieved if the 
multivector $\eta$ satisfies two conditions. The first is 
the `differential condition' 

\be
\begin{array}{ll}
\eta_{i_1\dots i_{n-1}\rho}\partial^\rho\eta_{j_1\dots j_n}-&
(\partial^\rho\eta_{i_1\dots i_{n-1} j_1})\eta_{\rho j_2\dots j_n}-
(\partial^\rho\eta_{i_1\dots i_{n-1} j_2})\eta_{j_1 \rho j_3\dots j_n}
-\dots\\[0.2cm]
&
-(\partial^\rho\eta_{i_1\dots i_{n-1} j_n})\eta_{j_1\dots j_{n-1}\rho}=0\quad,
\end{array}
\ee
which we shall write here in compact form as 
\be
\eta_{i_1\dots i_{n-1}\rho}\partial^\rho\eta_{j_1\dots j_n}
-{1\over (n-1)!}\epsilon^{l_1\dots l_n}_{j_1\dots j_n}
(\partial^\rho\eta_{i_1\dots i_{n-1} l_1})\eta_{\rho l_2\dots l_n}
=0 \quad.
\label{difcond}
\ee
The second condition, which follows from requiring that the terms with 
second derivatives of $f_1,\dots,f_{n-1}$ in the FI should vanish, is the 
`algebraic condition' 
\be
\Sigma + P(\Sigma)=0\quad, 
\label{alcond}
\ee
where $\Sigma$ is the tensor of order $2n$ given by the sum of $(n+1)$ terms
\be
\begin{array}{rl}
\Sigma_{i_1\dots i_n j_1\dots j_n} 
=&\eta_{i_1\dots i_{n}}\eta_{j_1\dots j_{n}} 
-\eta_{i_1\dots i_{n-1}j_1}\eta_{i_n j_2 \dots j_{n}}
-\eta_{i_1\dots i_{n-1}j_2}\eta_{j_1 i_n j_3\dots j_{n}}
\\[0.2cm]
-& \eta_{i_1\dots i_{n-1}j_3}\eta_{j_1 j_2 i_n j_4 \dots j_{n}}-
\dots
-\eta_{i_1\dots i_{n-1}j_n}\eta_{ j_1j_2 \dots j_{n-1} i_n} \;,
\end{array}
\label{Sigmadef}
\ee
and $P$ interchanges the indices $i_1$ and $j_1$ in $\Sigma$\footnote{
{}From the condition $\Sigma=0$ easily follows that in a $n$-dimensional space 
the (obviously decomposable) $n$-tensor 
$\eta_{i_1\dots i_n}=\epsilon_{i_1\dots i_n}$ 
defining the $\R^n$ volume element and the tensor 
$\eta_{i_1\dots i_{n-1}}(x)=\epsilon_{i_1\dots i_n} x^{i_n}$ are Nambu tensors 
\cite{ChaTak} \ie, satisfy the conditions (\ref{difcond}) and (\ref{alcond}).}.
Eq. (\ref{Sigmadef}) may we rewritten as
\be
\Sigma_{i_1\dots i_n j_1\dots j_n} = 
{1\over n!} \epsilon^{l_1\dots l_{n+1}}_{i_n j_1\dots j_n}
\eta_{i_1\dots i_{n-1} l_1}\eta_{l_2\dots l_{n+1}}\quad .
\label{ntensor}
\ee
Clearly, the algebraic condition is fulfilled if $\Sigma=0$. This implies in 
turn that the skewsymmetric tensor $\eta$ is decomposable
(\ie, it can be written as an exterior product of vector fields on $M$) and in 
fact, as conjectured in \cite{ChaTak}, 
it may be shown \cite{AG,GAUTH,HIE} that all N-P tensors ($n>2$) are 
decomposable (for $n=2$, eq. (\ref{alcond}) is trivial).

\medskip
\noindent
{\it b) Generalized Poisson structures}

Recently, another generalization \cite{APPB} of the ordinary PB has been 
proposed under the name of {\it generalized Poisson structures} (GPS) 
by extending the geometrical approach to standard Poisson 
structures \cite{Lich}. In these, a bivector $\Lambda\in\wedge^2(M)$ 
on a manifold $M$ defines a Poisson structure 
{\it iff} it has vanishing Schouten-Nijenhuis bracket (SNB) with itself, 
$[\Lambda,\Lambda]=0$. This condition, when generalized to multivectors 
of even order $\Lambda\in\wedge ^{2p}(M)$ provides the definition 
of the GPS (see below for the odd order case)
because for
\be
\Lambda = {1\over {(2p)!}}\,\omega _{j_1\ldots j_{2p}}\,\partial^{j_1}
\wedge \ldots \wedge \partial ^{j_{2p}}
\label{evenvector}
\ee
the requirement $[\Lambda,\Lambda]=0$ means that the coordinates of the 
{\it generalized Poisson} (GP) multivector $\Lambda$ satisfy the 
condition \cite{APPB}
\be
\epsilon^{j_1\dots j_{4p-1}}_{i_1\dots i_{4p-1}}
\omega _{j_1j_2\ldots j_{2p-1}k}\partial ^k
\omega_{j_{2p}\ldots j_{4p-1}}= 0\quad,
\label{genjaccond}
\ee
which is equivalent to the ($4p$--1)-point, $({4p-1 \atop 2p-1})$-terms 
{\it generalized Jacobi identity} (GJI)
\be
\epsilon^{j_1\dots j_{4p-1}}_{1\dots 4p-1}
\{f_{j_1},f_{j_2},\ldots ,f_{j_{2p-1}},\{f_{j_{2p}},\ldots ,f_{j_{4p-1}}\}\}=0
\label{genjacid}
\ee
where the {\it generalized Poisson bracket} (GPB) is also defined by
(\ref{nbrdef}) but for the $\Lambda$ in (\ref{evenvector}).
Notice that, as we shall see below, 
no further conditions are needed to remove the second derivatives 
from eq. (\ref{genjacid}), which is already free of them. 
As a result the $2p$-vector is constrained by the differential condition 
(\ref{genjaccond}) {\it only}.

The even GPS's have a clear differential geometrical 
origin due to their definition in terms of the SNB by the condition
$[\Lambda, \Lambda]$=0. Moreover, 
in the linear case one can find (an infinite number of) examples of 
even GPS defined by the Lie algebra cohomology cocycles \cite{APPB}. 
Indeed, for simple Lie algebras of rank $l$, there are $l$
antisymmetric tensors provided by the $l$
($2p_i$--1)-cocycles ($i=1,\dots,l$) \cite{AI} with coordinates 
${\Omega_{j_1\dots j_{2p_i-2}}}^\sigma_\cdot$, 
which define GP tensors of order ($2p_i-2$), 
\be
\omega_{j_1\dots j_{2p-2}}={\Omega_{j_1\dots j_{2p-2}}}^\sigma_\cdot x_\sigma
\quad ,
\label{cocycles}
\ee 
which satisfy (\ref{genjaccond}). These linear GPB's may 
be seen to be the analogues of
the even multibrackets defining higher order Lie algebras \cite{HIGHER} and, 
from this point of view, there is a one-to-one correspondence between these 
linear GPB and the higher order brackets of associative non-commuting 
operators. The time evolution, defined as in (\ref{tev}) but for
($2p-1$) Hamiltonians, is not a derivation of the GPB as it is in the
N-P structure.
In contrast with the N-P tensors, however, the GP $2p$-multivectors 
(\ref{evenvector}) are not decomposable in general because they do not need 
obeying the algebraic condition (\ref{alcond}).
It is easy to check, on the other hand, that any decomposable 
multivector of order $2p$, $p>1$, defines a GPS. 
Indeed, in this case 
$\Lambda$=$X_{1}\wedge\dots\wedge X_{2p}$ 
and using standard properties of the SNB [eq. (4.1) in the second ref. in
\cite{APPB}] it follows that
\be
\begin{array}{l}
[\Lambda,\Lambda]=
[X_1\wedge \dots\wedge X_{2p},X_1\wedge \dots\wedge X_{2p}]=
\\[0.2cm]
\qquad
\sum (-1)^{t+s} X_{1}\wedge\dots \widehat{X_s} \dots \wedge X_{2p},\wedge 
[X_s,X_t]\wedge
X_{1}\wedge\dots \widehat{X_t} \dots \wedge X_{2p}
=0
\end{array}
\label{DEC}
\ee
due to the appearance of repeated vector fields.

Much in the same way that on a Poisson manifold it is possible to 
define a Poisson cohomology \cite{Lich}, a GPB 
also defines a {\it generalized Poisson cohomology} \cite{APPB} through the 
Schouten-Nijenhuis bracket. 
Explicitly, if the $2p$-vector $\Lambda$ defines a GPS, the mapping
$\delta_{\Lambda}:\wedge^q(M)\to\wedge^{2p+q-1}(M)$ defined by 
\be
\delta_{\Lambda}:\alpha\mapsto[\Lambda,\alpha]\quad,
\label{GPcohomology}
\ee
is nilpotent since $[\Lambda,[\Lambda,\alpha]]=0$ and defines a 
$(2p-1)$-degree cohomology operator.

\medskip

Eq. (\ref{DEC}) and the decomposability of all N-P tensors show
that there is an overlap among the above generalizations of the standard
PS. This may be shown directly by noticing first that the GJI 
(\ref{genjacid}) is a full antisymmetrization of (\ref{fi})
\footnote{This fact was also known to L. Takhtajan (private communication).}. 
This observation leads to the following simple

\begin{lemma}
\label{lem1.1}
A N-P bracket (hence, satisfying the FI (\ref{fi})) verifies
\be
\epsilon^{j_1\dots j_{2n-1}}_{1\dots 2n-1}
\{f_{j_1},f_{j_2},\ldots ,f_{j_{n-1}},\{f_{j_{n}},\ldots ,f_{j_{2n-1}}\}\}=0
\quad.
\label{genjacid2}
\ee
\end{lemma}
\medskip
\noindent
{\it Proof:}\quad
Multiplying both hand sides of (\ref{fi}) by $\epsilon$ and using its 
antisymmetry, (\ref{fi}) is rewritten as
\be
\begin{array}{l}
\epsilon^{j_1\dots j_{2n-1}}_{1\dots 2n-1}
\{f_{j_1},f_{j_2},\ldots ,f_{j_{n-1}},\{f_{j_{n}},\ldots ,f_{j_{2n-1}}\}\}=
\\[0.2cm]
=n(-1)^{n-1}\epsilon^{j_1\dots j_{2n-1}}_{1\dots 2n-1}
\{f_{j_1},f_{j_2},\ldots ,f_{j_{n-1}},\{f_{j_{n}},\ldots 
,f_{j_{2n-1}}\}\}\quad;
\end{array}
\ee
hence, for $n\ge 2$, we obtain (\ref{genjacid2}), {\it q.e.d.} (for $n=2$ the 
N-P and the GPS reduce to the standard PS).

Eq. (\ref{genjacid2}), for $n=2p$, is the same as (\ref{genjacid}) and we 
conclude that 
{\it every Nambu-Poisson bracket of even order also defines a generalized 
Poisson bracket} \cite{mleon}.

\medskip
Due to the geometrical origin of the GJI condition, the GPS were
originally introduced \cite{APPB} for even order only:
the SNB of a $p\ (q)$-multivector $A\ (B)$ satisfies $[A,B]=(-1)^{pq}[B,A]$ 
and thus $[\Lambda,\Lambda]\equiv 0$ if $\Lambda$ is of odd order
(we are not including here the case of the `super' SNB \cite{AIPPB}).
Nevertheless, we may extend the GPS's by adopting the 
GJI (\ref{genjacid2}) for arbitrary (even or odd) $n$
as a first step in their definition.
In the odd case, the GJI is unrelated to the condition $[\Lambda,\Lambda]=0$
since it is trivially satisfied. 
But if we now define an odd-order GPB satisfying 
(\ref{genjacid2}) for $n$ {\it odd}, we find setting 
$f_i=x_i\,,\,i=1,\dots,2n-1$ that the coordinates of the 
associated $n$-vector $\Lambda$ must satisfy the differential condition 
(cf. (\ref{genjaccond}), (\ref{difcond})) 
\be
\epsilon^{j_1\dots j_{2n-1}}_{i_1\dots i_{2n-1}}
\omega _{j_1j_2\ldots j_{n-1}k}\partial ^k
\omega_{j_{n}\ldots j_{2n-1}}= 0\quad.
\label{genjaccond2}
\ee
For $n$ odd a second step now becomes necessary to cancel all second 
derivatives that appear in the GJI (\ref{genjacid2}). If we want to 
keep the GJI for odd-order brackets we have to impose an additional 
`algebraic condition' to the $n$-vector defining the structure.
Explicitly, this condition (for arbitrary $n$) is (cf. (\ref{alcond}))
\be
\epsilon^{i_1\dots i_{n-1}\,j_1\dots j_{n-1}}_{k_1\dots\dots \dots k_{2n-2}}
(\omega_{i_1\dots i_{n-1}\rho}\omega_{j_1\dots j_{n-1}\sigma}+
\omega_{i_1\dots i_{n-1}\sigma}\omega_{j_1\dots j_{n-1}\rho})=0\quad.
\label{secder}
\ee
For even $n$ this equation is automatically satisfied; this explains why 
there is no `algebraic condition' for even multivectors defining a GPS. 
In contrast, eq. (\ref{secder}) is an additional condition on 
$\omega$ for $n$ odd.

As a consequence of Lemma \ref{lem1.1},
conditions (\ref{genjaccond2}) and (\ref{secder}) 
must be extracted from conditions (\ref{difcond}) and (\ref{alcond}).
In fact, it is easily deduced that (\ref{genjaccond2}) follows (only) from 
(\ref{difcond}) and that (\ref{secder}) comes (only) from (\ref{alcond}).

Summarizing, extending the definition of GPS to odd brackets, 
the following general lemma follows

\begin{lemma}
\label{lem1.2}
The N-P tensors of even or odd order are a subclass of the 
multivectors defining the GPS, namely those 
for which the time evolution is a derivation of the bracket (or, in other 
words, the time evolution operator preserves the Poisson $n$-bracket 
structure).
\end{lemma}

We conclude this section by mentioning that one might think of using Lie 
algebra cocycles $\Omega_{i_1\dots i_{2p} \sigma}$ as the coordinates of 
a $(2p+1)$-vector $\Lambda$ leading to the odd bracket 
$\{f_{i_1},\dots,f_{i_{2p}},f_\sigma\}=
\Lambda(df_{i_1},\dots,df_{i_{2p}},df_\sigma)$ (see \cite{BMorr} for the 
trilinear case; cf. \cite{SV}).
However, although the differential condition for both the N-P (eq. 
(\ref{difcond}))
and odd GPS (eq. (\ref{genjaccond2})) are trivially satisfied for a constant 
multivector, 
this is not in general the case for the algebraic N-P (eq. (\ref{alcond})) and 
odd GPS (eq. (\ref{secder})) conditions. In contrast, any cocycle
defines an even linear GPS.

\section{Homology and cohomology}
\label{sec2}

We now compare the homological complexes underlying both 
structures [N-P, ($a$) and GPS, ($b$)].
First, let us recall the standard homology complex for a Lie 
algebra $\g$. The $n$-chains are $n$-vectors of $\wedge^n(\g)$ (for instance, 
left-invariant [LI] $n$-antisymmetric contravariant tensors on the associated 
group $G$, \ie, LI elements of $\wedge^n(T(G))\,$), and the homology operator 
$\partial C_n\to C_{n-1}$ is defined by 
\be
\partial(x_1\wedge\ldots\wedge x_n)=
\sum_{1\le l<k\le n}
(-1)^{l+k+1}[x_l,x_k]\wedge x_1\wedge\ldots\widehat x_l\ldots\widehat x_k\ldots
\wedge x_n \quad ,
\label{coder}
\ee
where $x\in\g$ and $[\ ,\ ]$ is the Lie bracket in $\g$; 
$\partial[\wedge^n(\g)]=0$ for $n\le 1$.
In particular, $\partial(x_1\wedge x_2)=[x_1,x_2]$ and, in this case, 
$\partial$ may be relabelled $\partial\equiv\partial_2$, 
$\partial_2:\wedge^n(\g)\to\wedge^{n-(2-1)}(\g)$.

\medskip
\noindent
{\it a1) Nambu-Lie homology}

Let us consider now 
a {\it Nambu-Lie} (N-L) {\it algebra} ${\cal V}$ of order $s$ in the 
sense of \cite{SP}\footnote{The case of the more 
general Nambu-Leibniz $s$-algebra 
(which does not assume the antisymmetry of the bracket \cite{LodPir}) is 
discussed in \cite{DT}. We thank L. Takhtajan for sending this 
paper to us.}. This means that there is a antisymmetric $s$-bracket 
$[\cdot ,\mathop{\dots}\limits^{s},\cdot ]:
{\cal V}\times\mathop{\cdots}\limits^{s}\times{\cal V}\to {\cal V}$, 
$[x_1,\dots,x_s]\in{\cal V}$ which satisfies the FI
\be
\begin{array}{l}
[x_1,\dots,x_{s-1},[y_1,\dots,y_s]]=
[[x_1,\dots,x_{s-1},y_1],y_2,\dots,y_s]
\\[0.2cm]
+[y_1,[x_1,\dots,x_{s-1},y_2],\dots,y_s]
+\dots+[y_1,\dots,y_{s-1},[x_1,\dots,x_{s-1},y_n]]
\end{array}
\label{fialgebra}
\ee
\ie, such that the map $[x_1,\dots, x_{s-1},\cdot ]:{\cal V}\to {\cal V}$ is a 
multiderivation of the N-L bracket.
The Nambu-Lie homology has been introduced by Takhtajan \cite{SP}.
Let $C_n$ be the $n$-chains 
$C_n={\cal V}\otimes \mathop{\cdots}\limits^{n(s-1)+1}\otimes{\cal V}$, 
$C_0={\cal V}$.
It is convenient to denote the arguments in the chains $C_n$ by
\be
(X_1,X_2,\dots,X_n,x)=
(x_{i^1_1},\dots,x_{i^1_{s-1}},x_{i^2_1},\dots,x_{i^2_{s-1}},\dots,x_{i^n_1},
\dots,x_{i^n_{s-1}},x)
\label{newchain}
\ee
where
$X_1=(x_{i^1_1},\dots,x_{i^1_{s-1}})\in {\cal V}^{s-1}$, etc. and 
$x\in {\cal V}$.
Now we consider the dot products $C_1\times C_1\to C_1$ and 
$C_1\times {\cal V} \to {\cal V}$ defined by
\be
X\cdot Y :=\sum_{i=1}^{n-1}
y_1\otimes\dots\otimes [x_1,\dots,x_{n-1},y_i]\otimes\dots y_{n-1} \quad ,
\label{dotprod1}
\ee

\be
X\cdot x :=[x_1,\dots,x_{n-1},x]
\quad.
\label{dotprod2}
\ee
Due to the FI (eq. (\ref{fialgebra})) these products satisfy
\be
X\cdot(Y\cdot Z)-(X\cdot Y)\cdot Z = Y\cdot(X\cdot Z)
\quad,\quad
X\cdot(Y\cdot z)-(X\cdot Y)\cdot z = Y\cdot(X\cdot z)
\quad.
\label{jaclike}
\ee
If these products were antisymmetric (\ref{jaclike}) would be the Jacobi
identity and thus, we would have defined a Lie algebra.
Although they are not, we can still define a Lie-type homology because 
the operator $\partial_s$ defined on $C_1$ by 
$\partial_s:C_1\to C_0 = {\cal V}$, 
$\partial_s:(x_1,\dots,x_s)\mapsto [x_1,\dots,x_s]$ and on 
$C_n$ by
\be
\begin{array}{rl}
\displaystyle
\partial_s(X_1,\dots,X_n,x)=&
\displaystyle
\sum\limits_{1\le i<j\le n} (-1)^{i+1}
(X_1,\dots,\widehat X_i,\dots, X_i\cdot X_j,\dots ,X_{n},x)
\\[0.2cm] &
\displaystyle
+\sum\limits_{1\le i\le n} (-1)^{i+1}
(X_1,\dots,\widehat X_i,\dots,X_{p+1},X_i\cdot x)\quad,
\end{array}
\label{ghom}
\ee
verifies\footnote{This is the case for the Leibniz algebras
\cite{LodPir} where we have a Lie-like homology in which the `bracket' is not
antisymmetric. The Jacobi-like conditions (\ref{jaclike}) 
ensure that $\partial_s$ is nilpotent.} $\partial_s^2=0$.
On $2$-chains, $\partial^2_s=0$ gives the `fundamental identity' 
which replaces the Jacobi identity for Nambu-Lie algebras.
For instance, for $s=4$ we have
$\partial_4 (x_1,x_2,x_3,x_4)=[x_1,x_2,x_3,x_4]$ and
$\partial_4^2$ on $C_2$ gives (cf. (\ref{fialgebra}))
\be
\begin{array}{l}
\partial^2 (x_1,x_2,x_3,x_4,x_5,x_6,x_7)=
[[x_1,x_2,x_3,x_4],x_5,x_6,x_7]+[x_4,[x_1,x_2,x_3,x_5],x_6,x_7]
\\[0.2cm]
+[x_4,x_5,[x_1,x_2,x_3,x_6],x_7]+[x_4,x_5,x_6,[x_1,x_2,x_3,x_7]]
-[x_1,x_2,x_3,[x_4,x_5,x_6,x_7]]
\end{array}
\label{d1}
\ee

\medskip\goodbreak
\noindent
{\it b1) GP-Lie homology}

Let us now look at the case of even GPS.
To this aim, consider a {\it higher-order Lie algebra} in the sense of 
\cite{HIGHER} (see also \cite{HW,GNE}) \ie, let $\g$ be a vector space endowed 
with an antisymmetric $s$-linear operation ($s$ even)
$[\cdot ,\mathop{\ldots}\limits^s,\cdot ]:
\g\otimes\mathop{\cdots}\limits^s\otimes\g\to\g$, 
which verifies the generalized Jacobi identity
\be
{1\over s!}{1\over (s-1)!}
\sum_{\sigma\in S_{2s-1}} (-1)^{\pi(\sigma)}
[[x_{\sigma(1)},\ldots,x_{\sigma(s)}],x_{\sigma(s+1)},\ldots,x_{\sigma(2s-1)}] 
=0\quad.
\label{NEWgenjacid}
\ee
In particular, if $s$ is even the $s$-bracket of associative operators 
defined by
\be
[x_{i_1},x_{i_2},\ldots,x_{i_s}]=
\sum_{\sigma\in S_s}(-1)^{\pi(\sigma)}
x_{i_{\sigma(1)}}x_{i_{\sigma(2)}}\ldots x_{i_{\sigma(s)}}
\quad.
\label{defmulti}
\ee
satisfies (\ref{NEWgenjacid})
(for $s$ odd, the $l.h.s$ in (\ref{NEWgenjacid}) is 
proportional to $[x_1,\dots,x_{2s-1}]$ rather than zero \cite{HIGHER}).

The $n$-chains are now elements of $\wedge^n(\g)$ and the homology operator
$\partial_s$ is the linear mapping 
$\partial_s:\wedge^n(\g)\to\wedge^{n-(s-1)}(\g)$ defined by

\be
\partial_s(x_1\wedge\ldots\wedge x_n)=
{1\over s!(n-s)!}\epsilon^{i_1\dots i_n}_{1\, \ldots\, n}
\partial_s(x_{i_1}\wedge\ldots\wedge x_{i_s})\wedge x_{i_{s+1}}\wedge\ldots 
\wedge x_{i_n}\quad.
\label{oldextended}
\ee
Denoting,
$\partial_s(x_{i_1},\dots,x_{i_s})=[x_{i_1},\dots,x_{i_s}]\in\wedge(\g)$
eq. (\ref{oldextended}) may be rewritten 
\be
\begin{array}{l}
\displaystyle
\partial_{s}(x_1\wedge\dots\wedge x_{n})=
\\[0.2cm]
\quad\displaystyle
\sum_{1\le i_1<\dots <i_s \le n}
(-1)^{i_1+\dots +i_s+s/2}
[x_{i_1},\dots,x_{i_s}]\wedge x_1\wedge \dots\wedge \widehat x_{i_1}\wedge
\dots\wedge
\widehat x_{i_s}\wedge\dots\wedge x_{n} \quad, 
\end{array}
\label{extended}
\ee
and the GJI may be also expressed as 
$\partial_s^2[\wedge^{2s-1}(\g)]=0$.
For instance, for $s=4$,
$\partial_4^2(x_{i_1}\wedge x_{i_2}\wedge
x_{i_3}\wedge x_{i_4}\wedge x_{i_5}\wedge x_{i_7})$ gives the GJI 
(eq. (\ref{NEWgenjacid})) which is the sum of ${7!/4!3!}=35$ 
terms 
\be
\begin{array}{l}
 [[ x_{i_1},x_{i_2},x_{i_3},x_{i_4}],x_{i_5},x_{i_6},x_{i_7}]
-[[ x_{i_1},x_{i_2},x_{i_3},x_{i_5}],x_{i_4},x_{i_6},x_{i_7}]
\\
+[[ x_{i_1},x_{i_2},x_{i_3},x_{i_6}],x_{i_4},x_{i_5},x_{i_7}]
-[[ x_{i_1},x_{i_2},x_{i_3},x_{i_7}],x_{i_4},x_{i_5},x_{i_6}]
\\
+[[ x_{i_1},x_{i_2},x_{i_4},x_{i_5}],x_{i_3},x_{i_6},x_{i_7}]
-[[ x_{i_1},x_{i_2},x_{i_4},x_{i_6}],x_{i_3},x_{i_5},x_{i_7}]
\\
+[[ x_{i_1},x_{i_2},x_{i_4},x_{i_7}],x_{i_3},x_{i_5},x_{i_6}]
+[[ x_{i_1},x_{i_2},x_{i_5},x_{i_6}],x_{i_3},x_{i_4},x_{i_7}]
\\
-[[ x_{i_1},x_{i_2},x_{i_5},x_{i_7}],x_{i_3},x_{i_4},x_{i_6}]
+[[ x_{i_1},x_{i_2},x_{i_6},x_{i_7}],x_{i_3},x_{i_4},x_{i_5}]
\\
-[[ x_{i_1},x_{i_3},x_{i_4},x_{i_5}],x_{i_2},x_{i_6},x_{i_7}]
+[[ x_{i_1},x_{i_3},x_{i_4},x_{i_6}],x_{i_2},x_{i_5},x_{i_7}]
\\
-[[ x_{i_1},x_{i_3},x_{i_4},x_{i_7}],x_{i_2},x_{i_5},x_{i_6}]
-[[ x_{i_1},x_{i_3},x_{i_5},x_{i_6}],x_{i_2},x_{i_4},x_{i_7}]
\\
+[[ x_{i_1},x_{i_3},x_{i_5},x_{i_7}],x_{i_2},x_{i_4},x_{i_6}]
-[[ x_{i_1},x_{i_3},x_{i_6},x_{i_7}],x_{i_2},x_{i_4},x_{i_5}]
\\
+[[ x_{i_1},x_{i_4},x_{i_5},x_{i_6}],x_{i_2},x_{i_3},x_{i_7}]
-[[ x_{i_1},x_{i_4},x_{i_5},x_{i_7}],x_{i_2},x_{i_3},x_{i_6}]
\\
+[[ x_{i_1},x_{i_4},x_{i_6},x_{i_7}],x_{i_2},x_{i_3},x_{i_5}]
-[[ x_{i_1},x_{i_5},x_{i_6},x_{i_7}],x_{i_2},x_{i_3},x_{i_4}]
\\
+[[ x_{i_2},x_{i_3},x_{i_4},x_{i_5}],x_{i_1},x_{i_6},x_{i_7}]
-[[ x_{i_2},x_{i_3},x_{i_4},x_{i_6}],x_{i_1},x_{i_5},x_{i_7}]
\\
+[[ x_{i_2},x_{i_3},x_{i_4},x_{i_7}],x_{i_1},x_{i_5},x_{i_6}]
+[[ x_{i_2},x_{i_3},x_{i_5},x_{i_6}],x_{i_1},x_{i_4},x_{i_7}]
\\
-[[ x_{i_2},x_{i_3},x_{i_5},x_{i_7}],x_{i_1},x_{i_4},x_{i_6}]
+[[ x_{i_2},x_{i_3},x_{i_6},x_{i_7}],x_{i_1},x_{i_4},x_{i_5}]
\\
-[[ x_{i_2},x_{i_4},x_{i_5},x_{i_6}],x_{i_1},x_{i_3},x_{i_7}]
+[[ x_{i_2},x_{i_4},x_{i_5},x_{i_7}],x_{i_1},x_{i_3},x_{i_6}]
\\
-[[ x_{i_2},x_{i_4},x_{i_6},x_{i_7}],x_{i_1},x_{i_3},x_{i_5}]
+[[ x_{i_2},x_{i_5},x_{i_6},x_{i_7}],x_{i_1},x_{i_3},x_{i_4}]
\\
+[[ x_{i_3},x_{i_4},x_{i_5},x_{i_6}],x_{i_1},x_{i_2},x_{i_7}]
-[[ x_{i_3},x_{i_4},x_{i_5},x_{i_7}],x_{i_1},x_{i_2},x_{i_6}]
\\
+[[ x_{i_3},x_{i_4},x_{i_6},x_{i_7}],x_{i_1},x_{i_2},x_{i_5}]
-[[ x_{i_3},x_{i_5},x_{i_6},x_{i_7}],x_{i_1},x_{i_2},x_{i_4}]
\\
+[[ x_{i_4},x_{i_5},x_{i_6},x_{i_7}],x_{i_1},x_{i_2},x_{i_3}]=0\quad.
\end{array}
\label{dcd1}
\ee
For the even linear GPS constructed from odd Lie algebra 
cocycles, the above GJI's truly reflect the underlying Lie algebra 
structure; this justifies the GP-{\it Lie} name given to this case.
These GJI are particular examples of those appearing in the strongly homotopy 
algebras \cite{LAST}, which contain `controlled' violations of the above
GJI  which may be introduced in our scheme by using a
suitable modification of the {\it complete BRST operator}
associated to ${\cal G}$ (\cite{HIGHER}, Theorem 5.2). 
These algebraic structures have been found relevant in closed string 
field theory (see the refs. quoted in \cite{LAST,HIGHER}).

\medskip
\noindent
{\it a2) Nambu-Lie cohomology}

Let us now consider the dual cohomology operations.
For the Nambu-Lie case we define $n$-cochains 
$C^n$ as mappings
$\alpha:{\cal V}\otimes\mathop{\dots}\limits^{n(s-1)+1}\otimes {\cal V}
\to {\cal A}$ 
where ${\cal A}$ is an abelian algebra (real field, for instance). 
Then, the cohomology operator 
$\delta_s:C^n\to C^{n+1}$ is defined as the dual of the homology operator 
$\partial_s$, 
$(C^n,\partial_s C_{n+1})=(\delta_s C^n, C_{n+1})$ where $(\ ,\ )$ denotes the 
natural pairing between chains and cochains.
Using this duality it follows immediately that the operator $\delta_s$ is 
defined ({\it cf.} \cite{GAUTH}) 
by its action on a cochain $\alpha\in C^p$ by
\be
\begin{array}{rl}
\displaystyle
(\delta_s \alpha)(X_1,\dots,X_{p+1},x)=&
\displaystyle
\sum\limits_{1\le i<j\le p+1} (-1)^{i+1}
\alpha(X_1,\dots,\widehat X_i,\dots, X_i\cdot X_j,\dots X_{p+1},x)
\\[0.3cm] 
\displaystyle
+
&
\displaystyle
\sum\limits_{1\le i\le p+1} (-1)^{i+1}
\alpha(X_1,\dots,\widehat X_i,\dots,X_{p+1},X_i\cdot x)\quad,
\end{array}
\label{gcoh}
\ee
where as in the homology case $X=(x_1,\dots,x_{s-1})\in {\cal V}^{s-1}$ 
and $x \in {\cal V}$. The proof that  $\delta^2_s=0$
is analogous to that for the Lie algebra 
cohomology coboundary operator if one thinks of $X_i\cdot X_j$ in 
(\ref{gcoh}) as a commutator, in which case eq. (\ref{jaclike}) looks 
like a Jacobi identity.

\medskip\goodbreak
\noindent
{\it b2) GP-Lie cohomology}

In the case of the linear GPS constructed on the dual of a Lie 
algebra we may introduce a cohomology operator dual to the homology one given 
in eq. (\ref{extended}). Acting on $n$-cochains $\alpha_{i_1\dots i_n}$
\be
\begin{array}{ll}
&
\displaystyle
(\delta_{s}\alpha)
(x_1,\dots,x_{n+s-1})=
\\[0.3cm]
&
\displaystyle
\sum_{1\le i_1<\dots <i_s \le n+s-1}
(-1)^{i_1+\dots + i_s+s/2}
\alpha([x_{i_1},\dots,x_{i_s}],x_1,\dots,\widehat x_{i_1},\dots, 
\widehat x_{i_s},\dots, x_{n+s-1})
\end{array}
\ee
or equivalently, setting 
$[x_{i_1},\dots,x_{i_s}]={\omega_{i_1\dots i_s}}^\rho x_\rho$ 
for definiteness,
\be
(\delta_s \alpha)_{i_1\dots i_{n+s-1}}={1\over s! (n-1)!}
\epsilon^{j_1\dots j_{n+s-1}}_{i_1\dots i_{n+s-1}}{\omega_{j_1\dots j_s}}^\rho
\alpha_{\rho j_{s+1}\dots j_{n+s-1}}\quad.
\label{incoord}
\ee
The nilpotency of $\delta_s$ follows from checking that \cite{APPB}
\be
\begin{array}{rl}
\displaystyle
(\delta^2_s \alpha)_{i_1\ldots i_{2s+n-2}}=&
\displaystyle
{s\over (s!)^2(n-1)!}
\epsilon^{j_1\ldots j_s k_1 \ldots k_{s+n-2}}_{i_1 \ldots i_{2s+n-2}}
{\omega_{j_1\ldots j_s}}^{\rho}
{\omega_{\rho k_1\ldots k_{s-1}}}^\sigma
\alpha_{\sigma k_{s}\ldots k_{s+n-2}}
\\[0.2cm]
\displaystyle
+ &
\displaystyle
{(n-1)\over (s!)^2(n-1)!}
\epsilon^{j_1\ldots j_s k_1 \ldots k_{s+n-2}}_{i_1 \ldots i_{2s+n-2}}
{\omega_{j_1\ldots j_s}}^{\rho}
{\omega_{k_1\ldots k_{s}}}^\sigma
\alpha_{\sigma \rho k_{s+1}\ldots k_{s+n-2}}=0\ ,
\end{array}
\ee
where the second term is zero since $s$ is even and the cochain $\alpha$ is 
antisymmetric 
in $(\rho,\sigma)$ and the first one is also zero since it encompasses the GJI.
Since this cohomology is based on multialgebra commutators, 
it applies to {\it linear} GPS.
For a general GPS, however, the operator (\ref{incoord}) is not defined, but 
the associated $2p$-vector $\Lambda$ still defines a generalized Poisson 
cohomology by (\ref{GPcohomology}).

\section{Concluding (physical) remarks}

\borrar{
\begin{table}
\renewcommand{\tabularxcolumn}[1]{>{\raggedright\arraybackslash}m{#1}}
\begin{center}
\begin{tabularx}{\linewidth}{|X|X|X|X|}
\hline
& N-P & GPS (even order) & GPS (odd order)
\\
\hline
Jacobi identity & Eq. (\ref{fi}) & Eq. (\ref{genjacid}) & 
Eq. (\ref{genjacid})
\\
\hline
Differential condition & Eq. (\ref{difcond}) & Eq. (\ref{genjaccond})
& Eq. (\ref{genjaccond2})
\\
\hline
Algebraic condition & Eq. (\ref{alcond}) & --- & Eq. (\ref{secder})
\\
\hline
Liouville theorem & Yes & Yes & Yes
\\
\hline
Poisson theorem & Yes & With restrictions & With restrictions
\\
\hline
Realization as associative operators & ??? & Yes & No
\\
\hline
\end{tabularx}
\caption{Comparison between Nambu-Poisson and generalized Poisson structures.}
\end{center}
\end{table}
}

The $n$-dimensional phase space of Nambu \cite{Na}
for the N-P structure associated with the volume element in $\R^ n$,
determined by an $n$-vector $x_i$,
has a divergenceless velocity field since, by (\ref{tev}),
\be
\partial^j \dot x_{j} = \partial^j \{H_1,\dots,H_{n-1},x_{j}\}=
\partial^j \epsilon_{i_1\dots i_{n-1} j }{\partial H_1\over \partial x_{i_1}}
\dots {\partial H_{n-1}\over \partial x_{i_{n-1}}}
=0\quad.
\ee
This analogue of the {\it Liouville theorem}
(a main motivation in Nambu's generalization of Hamiltonian dynamics) also
holds for the linear GPS given by the cocycles (\ref{cocycles}) since
$\omega_{i_1...i_{2m-2}}=x_\sigma {\Omega_{i_1...i_{2m-2}}}^\sigma_\cdot$
and ${\Omega_{i_1...i_{2m-2}}}^\sigma_\cdot$ is a constant antisymmetric
$(2m-1)$-tensor. Thus,
\be
\begin{array}{rl}
\displaystyle
   \partial^j{\dot x}_j= &\displaystyle
\partial^j(\omega_{i_1...i_{2m-3} j}\partial^{i_1}
 H_1...\partial^{i_{2m-3}}H_{2m-3})
\\[0.2cm]
\displaystyle
 =& \displaystyle 
\partial^{j}(x_\sigma {\Omega_{i_1...i_{2m-3} j}}^\sigma_\cdot)\partial^{i_1}
H_{1}...\partial^{i_{2m-3}}H_{2m-3}
=
{\Omega_{i_1...i_{2m-3}j}}^j_\cdot=0\quad .
\end{array}
\label{liou}
\ee
More generally, the conservation equation is clearly satisfied when the GPS on 
$M$ is defined by 
$\omega_{i_1...i_{2m-2}}=\partial^l{\tilde\omega}_{li_1...i_{2m-2}}$,  
and $\tilde\omega$ is an odd-order antisymmetric tensor \footnote{ 
The previous case of the linear GPS
is included here because one may take 
$
{\tilde\omega}_{li_1...i_{2m-2}}
=\frac{1}{(2m-2)!}
\frac{1}{n-2m+3}\epsilon^{jj_1...j_{2m-2}}_{li_1...i_{2m-2}}x_jx_\sigma
{\Omega^\sigma}_{j_1...j_{2m-2}}\ ,\ i,j,l=1,\dots,n
$
where $n$ is the dimension of $M$. Note that the second denominator in the 
last 
expression cannot vanish because the order of the bracket $(2m-2)$ never
exceeds the dimension $n$ of the manifold.}.

The Poisson theorem states that the PB of two integrals of motion is also an 
integral of motion. In N-P mechanics the extension of the Poisson theorem is 
guaranteed by the FI \cite{Ta} that may be rewritten as
\be
{d\over dt}\{g_1,\dots, g_n\}= \sum_{i=1}^{n} 
\{g_1,\dots,{d g_i\over dt},\dots, g_n\}\quad.
\ee
For the GPS, there is also an analogue of the 
Poisson theorem, although the
condition required for the constants of the motion ($g_1,\dots,g_k$) 
($k\ge 2p$) is more stringent.
Not only the $g$'s have to be constants of the motion, 
$\{g_i,H_1,\dots,H_{2p-1}\}=0\,,\,i=1,\dots,k$: the set
$(g_{i_1},\dots,g_{i_{2p-1}},H_1,\dots,H_{2p-1})$ of any $(2p-1)$ constants of 
the motion and the $(2p-1)$ Hamiltonians has to be in involution 
\ie, any subset of $2p$ elements has to have zero GPB 
(\cite{APPB}, Theorem 6.1).
This is because, in contrast with (\ref{fi}), where the $f_1,\dots
f_{n-1}$ may play the role of Hamiltonians, the GJI in (\ref{genjacid}) 
includes GPB's which contain Hamiltonians and more than one constant of the 
motion.

\medskip
We would like to conclude with a comment concerning quantization. As pointed 
out by Nambu himself \cite{Na}, the antisymmetry property is necessary to 
have Hamiltonians that are constants of the motion in Nambu's mechanics.
This also applies to the higher-order N-P structures \cite{Ta}, and remains 
true as well of the GPS in \cite{APPB}.
The structure of the FI makes the N-P bracket in \cite{Ta}
specially suitable for the 
differential equation describing the time evolution of a dynamical quantity.
Nevertheless, the standard quantization of Nambu mechanics is an open problem 
likely without solution (see \cite{SV} in this respect).
There is a simple argument against an elementary quantization of N-P mechanics 
in which one tries to keep the standard one-to-one correspondence among certain 
dynamical quantities, their associated quantum operators and the infinitesimal 
generators of the invariance groups. It is physically natural to 
assume that these quantum operators, $x_i$ say, are associative.
But if so, it is not difficult to check \cite{HIGHER} that any commutator 
$[x_1,\dots,x_s]$ defined by the antisymmetrized sum of their 
products, as in (\ref{defmulti}), does not satisfy the FI. 
For odd $s$-brackets, moreover, we find (\cite{HIGHER}, Lemma 2.1) that
it is not possible to realize the GJI in terms of odd multibrackets,
since then the $r.h.s.$ of eq. 
(\ref{NEWgenjacid}) is replaced by $[x_1,\dots,x_{2s-1}]$.
Thus, for the odd case (which includes Nambu's) a multibracket of associative 
operators defined as in (\ref{defmulti}) leads to an identity which is 
{\it outside} the original N-P algebraic structure. 
For $s$ even, however, eq. (\ref{NEWgenjacid}) holds.
The resulting identity, however, is not the FI, but the GJI associated with 
the GPS introduced in \cite{APPB}. Thus, a natural correspondence between 
multibrackets and higher order PB exists only for the even multibrackets 
and the GPS's.
The associativity of the quantum operators is not compatible with the 
derivation property of the N-P bracket which leads to the FI (\ref{fi}).
Such a compatibility exists for the even GPS; however, in this case the time 
evolution fails to be a derivation of the GPB making it more difficult 
to establish a dynamics already at the classical level.

The above discussion indicates that, in Nambu's words, `quantum theory is 
pretty unique although its classical analog may not be'. 
The quantization of higher-order Poisson brackets requires renouncing to some 
of the standard steps towards quantum mechanics\footnote{A formal 
quantization of the $n$=3 Nambu mechanics case, where the 
Nambu bracket itself is replaced by an $\hbar$-deformed one, 
has been performed in \cite{DFST}. For the `quantization deformation'
and $*$-products, see the references quoted in \cite{DFST} and 
also \cite{BER}.}. 
But it may well be (see also \cite{MS}) that classical {\it mechanics} is 
pretty unique too if the term `dynamical system' is restricted to its physical 
(rather than mathematical) meaning.

\medskip
\noindent
{\bf Acknowledgements.}
The authors wish to thank L. Takhtajan for his comments on the manuscript.
This research has been partially supported by the CICYT and the DGICYT, Spain
(AEN 96-1669, PR 95-439). 
J.M.I thanks the EU (HCM programme) for financial support.
J.A. and J.C.P.B. wish to acknowledge the kind hospitality extended to them
at DAMTP and J.C.P.B. wishes to thank the Spanish Ministry of Education and 
Culture and the CSIC for an FPI grant.


\end{document}